\documentclass[oneside,twocolumn]{article}
\usepackage[utf8]{inputenc}
\usepackage{nopageno}
\usepackage{url}
\usepackage[backend=biber,style=oscola]{biblatex}
\usepackage{xcolor}
\usepackage{graphicx}
\usepackage{enumerate}
\usepackage{array}
\usepackage{ragged2e}
\usepackage{amsmath}
\usepackage{appendix}
\usepackage{authblk}
\usepackage{mathtools}
\usepackage{mathptmx}
\usepackage{mathrsfs}
\usepackage{amssymb}
\usepackage{amsmath}
\usepackage{amssymb}
\usepackage{csquotes}
\usepackage{dirtytalk}
\usepackage{hyperref}
\usepackage{dirtytalk}
\usepackage{lipsum}
\title{The Dangers of Computational Law and Cybersecurity; Perspectives from Engineering and the AI Act}

\author[1]{Kaspar Rosager Ludvigsen}
\author[2]{Shishir Nagaraja} 
\author[3]{Angela Daly}

\affil[1]{Department of Computer and Information Sciences, University of Strathclyde, kaspar.rosager-ludvigsen@strath.ac.uk}
\affil[2]{Department of Computer and Information Sciences, University of Strathclyde, shishir.nagaraja@strath.ac.uk}
\affil[3]{Leverhulme Research Centre for Forensic Science and Dundee Law School, adaly001@dundee.ac.uk}

\date{June 2022}

\begin{document}

\twocolumn[
  \begin{@twocolumnfalse}

\begin{abstract}


Computational Law has begun taking the role in society which has been predicted for some time. Automated decision-making and systems which assist users are now used in various jurisdictions, but with this maturity come certain caveats. Computational Law exists on the platforms which enable it, in this case digital systems, which means that it inherits the same flaws. Cybersecurity is one framework which addresses these potential weaknesses, and in this paper we go through known issues and discuss them in the various levels, from design to the physical realm. We also look at machine-learning specific adversarial problems, which entail further weaknesses. Additionally, we make certain considerations regarding computational law and existing and future legislation. Finally, we present three recommendations which are necessary for computational law to function globally, and which follow ideas in safety and security engineering. As indicated, we find that computational law must seriously consider that not only does it face the same risks as other types of software and computer systems, but that failures within it may cause financial or physical damage, as well as injustice. The consequences of Computational Legal systems failing are in this sense greater than if they were merely software and hardware. And if the system employs machine-learning, it must take note of the very specific dangers which this brings, of which data poisoning is the classic example. Computational law must also be explicitly legislated for, which we show is not the case currently in the EU, and this is also true for the cybersecurity aspects that will be relevant to it. But there is great hope in EU's proposed AI Act, which makes an important attempt at taking the specific problems which Computational Law bring into the legal sphere. Lastly, our recommendations for Computational Law and Cybersecurity are: Accommodation of threats, adequate use, and that humans must remain in the centre of their deployment. The latter is primarily for the abilities humans process and which allow them to handle emergencies.
    
\end{abstract}

  \end{@twocolumnfalse}
]

\maketitle

\section{Introduction}
Despite law being established as an academic discipline for a significant period of time, it is still rather undefined and lacks the rigour that other disciplines possess. There are ongoing debates and unresolved questions as to when law is deductive or inductive\footnote{And while these are vital, we cannot discuss them further in this paper. For a contrary opinion, see footnote 4, page 98 in \cite{Gaudencio2022}.}, so for computational law (CL) to even claim being future-proof misses the mark. Like the existence of exotic matter in astronomy, you may deduce or assume their existence, but empirical evidence will eventually prevail and show whether it is worthwhile\footnote{Past questions on this are still not fully answered, see \cite{Malaney1988}.}. The same can be said for CL, as its implementation should not be forced because of powers outside the academic and legal sphere\footcite[10 - 11]{Hoffmann-Riem2021}, or for the sake of profit. Increased use must be justified, regardless of whether some types of technology are forced upon everyone without any other reason than power\footnote{See any work on facial recognition or contract tracing, e.g., \cite{Rezende2020, Siatitsa2020, White2021}.}.

CL can be defined by its usage, like many other fields of law\footnote{See ongoing work on how this is understood, like \cite{Schafer2022}.}. Law in general, and CL too, contain an important modifier: 

They are affected by the technological systems they exist in\footnote{There are diverging opinions on this, see \cite{Koops2006} versus \cite{Ohm2010}.}, and CL expresses this in a much more extreme manner than any other field. It would not even exist without its extra-legal roots\footnote{See, e.g., \cite{Kelso1946, DanielHillis1982}.}. Computer science provides the basis and logic for CL, and in turn brings the same drawbacks\footnote{While this is not a new or unknown phenomena, it is worth discussing and elaborating on, both to make it clearer for policymakers, users and the public in general.}.

A good parallel to describe this, would be the event of IoT devices\footnote{IoT devices can be part of a CL system, e.g., the data could feed into a system which decides on the basis of it.}. The increased adoption of these brings in a new danger at every step, as all of them are (usually) connected to a network and make use of software. Every danger posed to computers and digital systems therefore exist in these devices, which are often used close to humans or for critical infrastructure, increasing the amount of possible accidents and the risk of damage\footnote{See any literature on this problem, \cite{Ludvigsen2022, Fu2020, Rizvi2018}.}.

CL shares this, every logical or otherwise systemic pitfall can pose a danger. For example, circumvention and adversarial law usually relies on a system that consists of courts and subjects seeking to abuse it\footnote{Extra-legal means to circumvent or abuse the legal system exist too, very much analogously to how many different ways an adversary can attack digital systems.}. Doing the same in CL only requires obfuscating the actions within the logic of the code or otherwise abusing technical limitations or faults of the system. The different types of parties which can do this to these systems are many times larger, than the sum of money and influence required to do this within pre-existing legal systems.

This brings us to some unique points about what kind of vulnerabilities CL will always have. Cybersecurity (which now applies to CL) consists of certain assumptions, either expressed directly or seen but not necessarily explicitly mentioned by those who practice it\footnote{We will not provide further argumentation for the assumptions, as these are not stringently codified in cybersecurity at the time of writing, but matter to how it is deployed in practice. For example, zero-day exploits may be dangerous, but are something you must prepare for and no one expects everything to be perfectly preventable.}. One of these is the assumption that there will never be a perfect defence. We are never reaching an equilibrium of defences and attacks, and the risk of incursions that can succeed will always exist. A major consequence of CL either becoming or already being implemented into national legal systems, is that they can now suffer adversarial failures (from adversarial attacks). Therefore, ironically, CL as a tool for further automation may be a target for automated adversarial attacks and loophole analysis itself.

Like in law, adversary merely refers to the subject being against the target, but the attacks and failures entail weaknesses which only increase in sophistication and consequences as the systems or technology used becomes more complicated and increasingly used. To illustrate this, imagine the difference between a court system which only partially or does not rely on CL, and one which does. The former cannot be brought to a halt or fully hijacked by an adversary. But a court system which makes fully or mostly use of CL definitely can, unless it has backups and redundancies that enables it to revert to a non-CL state.

In this sense, CL inherently makes a legal system more vulnerable, in turn potentially damaging rights, individuals, corporations and even the state itself. It is from here its relationship with cybersecurity must be scrutinised, understood and realised, which we can only discuss so much here.

In this paper, we take a very narrow look at one case, which is adversarial attacks on machine-learning (ML) models and any system that makes use of them. Computational legal systems that consist partially or fully of these, in any shape or form, will be vulnerable to attacks on the classifiers\footcite{Fredrikson2015}, one of the fundamental parts of these systems, which all the learning is used on and to form, and attacks to extract the data\footcite{Tramer2016} which they are created on, or the source code of the entire system. These are all well known, but for CL, they create barriers which make the practical implementation of these systems potentially unsafe.

Safety here refers to accidents or losses, while cybersecurity refers to lowering the risk of adversarial failures. It is said that safety is increasingly covering cybersecurity\footnote{Acknowledged by the European Union, see page 10 of ‘MDCG 2019-16 Guidance on Cybersecurity for medical devices’, \url{<https://ec.europa.eu/docsroom/documents/41863>}, last accessed 30 June 2022.}, as seen with the IoT example above, and this implies that CL must develop defences in its technical implementation and logic. Limiting itself to ways where risks can be mitigated or hazards can be controlled is suitable, and follow the tradition of anything that absorbs cybersecurity as part of its being\footnote{Ibid. But any type of product which gains network connectivity could be a good example. For a general overview in the EU, see \cite{Banasinski2021}.}

Section 2 defines what we see as CL and cybersecurity. After this, we discuss the weaknesses which CL contains in 5 levels. In Section 3, we dive further into machine-learning specific issues with CL, and we also comment on the problems which black boxes and authentication bring. Section 4 contains a brief legal commentary on how cybersecurity is regulated in the EU, with a focus on its effect on CL, and how the EU's future AI Act\footnote{Proposal for a Regulation of the European Parliament and of the Council laying Down Harmonised
Rules on Artificial Intelligence (Artificial Intelligence Act) and Amending Certain Union Legislative Acts,
COM/2021/206.} may affect CL as a field. We then provide general recommendations for the future, because of the obvious consequences which CL and cybersecurity together bring in Section 5, Section 6 has some ideas for future work, and finally, Section 7 concludes the paper.



\section{Computational Law and Cybersecurity}

CL and cybersecurity naturally fit together, because the former interacts or exists in some digital space which requires software and/or hardware. This section combines the two and looks at weaknesses.

\subsection{Definitions}

CL can be defined as:

\say{\emph{The techniques of computational logic, applied to the semantic rules as well as the data, form the basis of a computational law system.\footcite[206]{Love2005}}}

CL is considered to be automation of legal reasoning. But from the quote by Love, we can see that there are essentially two broad definitions, either purely automation or broadly applying computational logic to law. We will consider both in this paper.


\subsubsection{Forms of Computational Law}

To better contextualise CL, we mention selected forms here. This is not an exhaustive list, and some applications will cover several areas.

\begin{itemize}
    \item Automated assistance within legal systems. TurboTax\footnote{See \cite{Marr2012}, for an example of function and issues.} (while not developed by a State) is always brought up, but the Danish state is another example where income and other relevant information is automatically sought and filled in\footnote{\url{https://lifeindenmark.borger.dk/economy-and-tax/the-danish-tax-system/a-general-introduction-to-the-danish-tax-system}, last accessed 30 June 2022.} in tax returns and similar forms, leading to automated experiences for most subjects. On the back-end, this type of CL is also used by the authorities. 
    
    \item Contract related fields, like smart contracts\footcite{Dixit2022}, as well as other types or derivatives, are considered CL, but are often found in their own sub genres in relation to the technology which they exist in. These see primarily use in private law\footnote{However, even public authorities and states will be private law subjects to the companies they create contractual relations with.}, and will not always be able to execute their contents fully without assistance from litigation or other means\footnote{For an elaboration of these issues, see \cite{Pitschas2011, DiAngelo2019, Werbach2017}.}. The latter comes down to needing state or legal system assistance to fulfill them, in case of disputes or disagreements.
    
    \item Automated legal decision-making. This can range from public legal decisions on subsidies, pension, automated payments\footnote{The wrongful texts on Estonian AI should have referred to the development of automated systems for certain types of payment, see \url{https://www.just.ee/en/news/estonia-does-not-develop-ai-judge}, last accessed 30 June 2022.} or company registration, with a much bigger potential in the future\footnote{For an overview in public legal systems, see \cite{Roehl2022}.}. For future applications, the research community and the industry as such tends to focus on automated tools or AI judges\footnote{Some examples could be \cite{Morison2019, Wang2020a, Muhlenbach2020}.} but this seems very far-fetched with the capabilities of current CL systems.
    
    \item Quantum CL, if the event of quantum computing occurs, should be its own separate field \footcite[305]{Atik2021}. Primarily due to the power imbalance, but also the new physical constraints of the system. The former refers to the substantially increased amount of computing power these will have over conventional hardware, the latter to the real tangible physical difference which quantum computers contain, meaning that new threat models and measures will have to be developed\footnote{There is a wealth of post-quantum research which is ongoing, see, e.g., \cite{Ahn2022}.}.

\end{itemize}

What these all have in common, is that they exist in or as software and make use of hardware, while containing a level of automation and operate within or as the legal system. This entails that cybersecurity is an issue for all of them.

\subsubsection{Cybersecurity}

Cybersecurity can be defined as freedom from adversarial failures\footcite{Leveson1995}, but there exists other definitions outside of traditional security engineering. We chose this because it is narrow and focused on mitigating or limiting the damage of what failure to defend against attacks may cause.

As mentioned, this freedom is hard to attain, and it is therefore more of a goal to strive for, than a goal to reach\footnote{And can be expanded to include the process of attaining security, see \cite[17 - 18]{Radoniewicz2022}.}. Adversaries successfully attacking a system does not always lead to safety failures, but with the few examples we have mentioned, there is a risk they will. Safety is the general idea of freedom from accidents or losses\footcite{Leveson1995}, where both financial as well as human losses are included.

Cybersecurity itself is very diverse, and includes everything from encryption, good practice for building databases\footcite{Ibrahim2020} and naming conventions\footcite[259-271]{Anderson2020}, to physical elements like air gaps\footnote{Physical gaps between the system and the internet or just public space. Guri has written a range of papers on how to mitigate and understand air gaps, see e.g., \cite{Guri2021}.}, or access control\footcite{Valenzano2014} to both the servers or robots\footnote{Each area could be analysed specifically with regards to CL, but this is not the paper for that.}. 

What we focus on is the additional layer of complexity and potential risk which cybersecurity adds to CL. Very much like safety engineering with the event of the industrial revolution, cybersecurity needs to be integrated and considered when designing and also deploying CL systems. Gone are the days where the application and idea behind CL is purely discussed, as there are weaknesses which certain types of logic or choices add, giving rise to practical impacts.


\subsection{Weaknesses}\label{sectionweakness}

CL can exist in a theoretical manner without considering cybersecurity, but when deployed in practice, defences are needed in various ways. Stochastic or completely unpredictable issues are not considered because of their extraordinary nature. But the limitations to what can be considered special enough to warrant lack of liability depends on the jurisdiction, to this, see our past work\footcite{Ludvigsen2022}, for an analysis combining private law and cybersecurity within a Danish context.

\subsubsection{General Issues}


With inspiration from existing literature, we sketch an overarching conceptual understanding of how cybersecurity matters to CL.

Weakness in CL can be described in the following levels:

\begin{enumerate}
    \item Logic or design.
    \item Software.
    \item Hardware.
    \item System or network.
    \item Physical.
\end{enumerate}

If the logic or the design of the system does not consider most common threats, and has not been thoroughly tested, cybersecurity adds a layer of weakness to all levels. Testing for hardware is the same as software, but level 4 is different.

Broadly, all the attacks and defences work on individual levels (each substation), but if these are not applied uniformly, the weakness will consist of other systems (not those related to the CL), because individual weaknesses will impose a risk on everyone in the system, making level 4 exceptionally difficult. This includes the very network(s) that CL use. 

Atttacks on the physical layer, level 5, is to usually enable an attack on a lower level, except for those unique to it such as physical destruction, but shoulder-surfing\footnote{Refers to snooping, observing literally over a shoulder or otherwise, to deduce or directly observe passwords or other information. There exists plenty of research on the matter, see, e.g., \cite{bace2022}.} or physical side-channel attacks\footnote{Physical, unlike digital side-channel attacks, refers to readings which then allow an adversary to manipulate or otherwise know something they should not. This could be electromagnetic and so on, see following article for a great overview with a focus on neural networks, \cite{Real2021}.} are examples of the first.

Most levels are described or specified better in existing literature. Regardless of this, there are some unique additional explanations needed for each.

\begin{figure*}[h] \label{weakness}
    \centering
    \caption{Visualisation of the 5 weakness levels and their interactions.}
    \includegraphics[width=\textwidth]{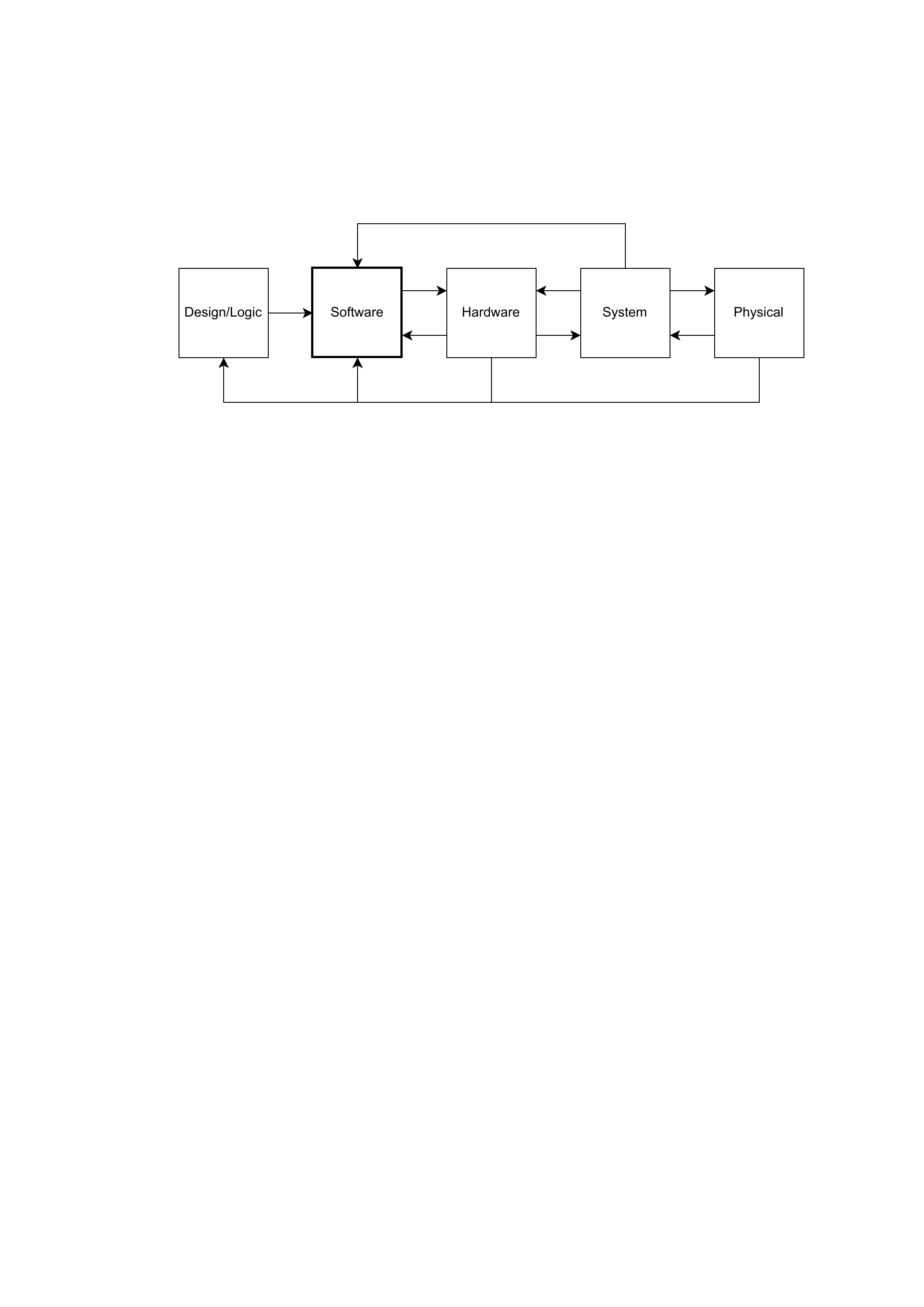}
    
\end{figure*}

To illustrate the division, we include an \hyperref[weakness]{visualisation} which shows the interactions between the different levels.

\subsubsection{Logic}

Cybersecurity confer devastating and potentially permanent weaknesses to CL systems through its choice of logic, as it is what defines what should be understood and done. This is explained by how it is constructed, e.g., through the use of machine-learning or expert decision-making, or through choices of programming language or other technical design decisions. These must be not only uncovered and possibly mitigated, but some may warrant the removal of the system entirely, if any of the core weaknesses compromise any basic cybersecurity attributes at large. For example, machine-learning based systems may contain weaknesses that allow both data and the code behind it to be easily retrievable, and if defences against this are not vigilantly updated and improved throughout the life-cycle of the system, they should not even be used in the first place.

Another could be the logic behind what constitutes a decision in the system\footnote{This applies to all types of platforms CL can use.}, becoming law, which may be circumventable or possible to be hijacked, either through known techniques or some discovered by fuzzing\footnote{Fuzzing is the practice of testing arbitrary or deliberate inputs or actions, and seeing how the system reacts to it. Some that can cause adversarial attacks may be discovered at the design stage, others later. For an overview, see \cite{Mcnally2012}.}. Circumvention or hijacking can be seen as obfuscation or abuse by known constraints of the system, which is known in law as: 

Loopholes\footnote{There exists an ocean of research on the concept, good specialised examples could be \cite{espionosa2022, Butler2022}.}, deliberate non-compliance without enforcement\footnote{For an empirical study with interesting perspectives on this, see \cite{Shodunke2022}.} or stalling for time\footcite{Mazur2022} through litigation or in public law\footnote{This list in not exhaustive, but represents common areas.}. CL allows circumvention etc., that does not necessarily require humans or much effort measured by time, which increases the risk as to whether it will be used\footnote{Cheap externalities usually leads to increased use, as we have seen with the data sharing, see, e.g., \cite{Ichihashi2021}. Worth noting that even if elements make higher profits for shareholders, this does not mean decreased costs for users or better rights, see \cite{Deibert2022}. This analogy persists with CL as well, as it will end up being extremely cheap to attack and abuse the systems going forward.}.


\subsubsection{Software}

CL is expressed in and usually executes within the software level. Pandora's box is therefore opened in regards to weaknesses, anything that is possible with the very software\footnote{This could be the AI that decides, or one of several subsystems.} that is or the CL resides in, expresses the opportunities of attack and failures. This could be in the form of denial-of-service of various kinds, leading to loss of integrity or availability. The latter is especially important if the CL is the only source of decisions on for example legal subsidies or permits, and if such a system is taken out, and there are no redundancies, systemic financial loss is very possible\footnote{As is mental damage to those who are denied finances to live from, as could be the case with pension or other types of social services.}. Software is usually also the target of attacks from the other levels, and attacking the CL software is very possible through other software in the system somewhere, which we focus on further below.

\subsubsection{Hardware}

Attacking software through hardware is classic, as seen with the Spectre attack on CPUs\footnote{There exist many good papers on defences and solutions against it, see e.g., \cite{Kadir2019}.}, and recently Hertzbleed\footcite{Wang2022}, allow for various actions going from hardware to software. In a CL context, this would allow attackers everything from stealing data, to escalation attacks that would give them control of the software, or simply a way to destroy it through ransomware attacks with no way to reverse the encryption of files.

Defences against this are often physical, but as attacks like Spook.js\footcite{Agarwal2022} show, you do not even need to have such access to perform the attack. In this sense, hardware only adds to the burden of defending CL systems, and the most important detail of all here, is there is no way to prevent all side or covert channels from being exploited, as they often are caused by deliberate decisions in hardware architecture. This is best seen with the Spectre attack and its derivatives, as the weakness that allows it also increases CPU performance immensely, and will therefore not (unless by law) be changed or removed. Analogies to this will exist in many types of hardware.

\subsubsection{System}

As mentioned above, attacks from other types of software towards the one which the CL system resides in is the biggest threat overall. Spook.js\footcite{Agarwal2022} does this partially too, as it enables attacks in the browser, which in a CL context could result in an adversary stealing information from singular users accessing CL decisions or data which is being provided to such a a system. On the provider end, operation systems, proprietary software or even malicious antivirus\footcite{Prakash2016} or just plain malware, can all cause one or several types of software or potentially hardware to be compromised. In this context, the weakness is the entire infrastructure where we rely on many types of software and hardware at once.

Solutions like trusted or trustworthy\footnote{Trustworthy must be made to never fail, whereas trusted merely is the idea that it should not fail, but there is no assurance it never will, \cite[13]{Anderson2020}.} hardware are not enough, and attackers merely need to exploit or use well known techniques to gain some type of access, then lie in wait until they can use their position to potentially reach the system which houses the CL. Even from the perspective of the user, weaknesses on their side may allow adversaries to manipulate or violate the confidentiality of CL system, adding an additional attack venue. In this sense, the system level weakness of CL is most likely the most severe, as the potential battlefield of different options and tools which adversaries can use are frankly many times greater than any of the others. 

\subsubsection{Physical} 

Destroying servers physically, or abusing physical interfaces such as USB\footcite{Thomas2021} all constitute primary reasons to have backups and redundancies for CL systems. Physical attacks, unless aimed at destruction, will be means to escalate and gain access to hardware and software. It may seem like the simplest area, but defending and perhaps expecting employees to abuse their knowledge of location of servers and so on, is in its own way paramount in cybersecurity regarding CL systems. 

As mentioned, physical side-channel attacks are another matter where defences must be erected, but many types will regrettably always exist due to their passive or irreplaceable nature\footnote{Electromagnetic emissions are close to impossible to prevent efficiently in all systems, which means this can usually always be deployed. For insights into this, see \cite{Sayakkara2021, Sayakkara2019}. It could also be audio, anything that can reveal the inner workings of hardware is possible to make use of.}, either due to the costs for preventing any leakage or disabling designs which are necessary for the functioning of the hardware.

\subsection{Strengths}

On the contrary, CL does contain certain strengths over non-computational systems. Distributed and massive decision-making\footnote{On the contrary, see \cite{Cobbe2019a}.}, simplified and efficient support and search powers (to help users or citizens) and the other classic benefits of automation apply\footnote{While this is contentious, and requires more empirical research to confirm, there are indications towards this direction, see \cite{Zalnieriute2019}.}. However, all of it requires the right kind of humans in-the-loop and legislative framework. Assuming the latter is true in a system, the strengths of CL are quite clear. Regardless of this, the strengths may not outweigh the costs, which we will comment on \hyperref[recommendations]{later}.

\section{Cases}

To specifically illustrate the interaction between CL and cybersecurity, we take a look at some examples.

\subsection{Machine-Learning Specific Issues}\label{ml}

Existing taxonomies and threat models for machine-learning (ML) models and systems sufficiently describe the problem \footnote{See examples like \cite{Gupta2020, Papernot2018}. For a practical summary of the current threat model landscape, see \url{https://docs.microsoft.com/en-us/security/engineering/threat-modeling-aiml}, last accessed 30 June 2022.}. But in the context of CL, we can deliberately go past the CIA triad, confidentiality, integrity and availability\footnote{Failures in each may overlap during specific successful attacks. The CIA triad is a core concept in cybersecurity at large, and is understood in a literal sense.}, and focus on:

\begin{enumerate}
    \item Poisoning (modification) or possession of the training data.
    \item Attacks on the model or during the creation of it.
    \item Hijacking of the communication from or to the model or where it operates.
\end{enumerate}

\subsubsection{Poisoning or possession of data}

For ML based CL systems, this is where the biggest risk lies. Not only can all training data, personal or not, be stolen and published or used to make a competitive model, but these can also result in consequences which cannot be seen until the system has made an unfair or outright dangerous decision. Even if it did not matter whether the SyRi system was ML based or not\footcite{Rachovitsa2022}, an attack with similar or more severe consequences like in the Post Office case\footnote{\cite{Christie2020}. The loss of lives were a later consequence, but are still significant.}, could be caused by deliberately changing the data to make a system commit to wrong decisions or actions. There are defences, but vigilance by every human involved, combined with an understanding that data may be poisoned in the first place, and regular auditing, must be prioritised to minimize the threat which this constitutes.

\subsubsection{Model Attacks}

Not unlike other types of software, protecting the model at the design stage is vital, as adversaries can inject or modify parts which may, like above, cause decisions with very legal or physical consequences due to its CL nature\footnote{Or simply steal the model and use it for themselves.}. Unlike poisoning, model attacks may be used by journalists and competitors to attempt to reveal the contents of the model for various purposes. Activism in cybersecurity is well known\footnote{There exists plenty of research on this, from many different types of sciences, e.g., \cite{George2018}.}, and may play a role here and elsewhere, and with the position CL takes in society, this clash may be further exacerbated. Like other types of activism, which include activities that may be considered criminal, this is typically divided into black\footcite{Kwon2018}, white\footcite{Schrock2016} and grey hatted\footcite{Thomas2019}, but what matters the most, is the danger that each possess. This must be considered before using CL with ML, as the adversaries may cause undesirable outcomes for the provider, user and citizen respectively, explicitly by revealing the model illegally, ethically or unethically.

\subsubsection{Communication Attacks}

While less dangerous than the two others, and seen heavily elsewhere in various ways, communication attacks, if done in a CL system with multiple steps, could potentially cause some of the same issues as above. Traditionally, these are divided into several categories, such as replay, hijacking or modification. The central part is that what is sent and received is attacked, not necessarily the software itself, in this sense its actions\footnote{Papers which give an overview on this are, e.g., \cite{Syverson1994, Xenofontos2022, Huitsing2008}.}. Contents of the communication could lead to changes in decisions, or the system could just stop functioning from wrong received input. 


\subsection{Black Boxes}

Black boxes may exist deliberately, or for political or unknown reasons, but from a cybersecurity perspective, they are not wanted or useful. Trade-secrets or patents can warrant this, but the general idea is strong security, not secrecy\footcite{Kerckhoffs1883}. CL suffers further from this through the loss of legitimacy. Transparency and openness about what a system can and should do will increase trust and confidence and vice versa. The exceptions for this would be surveillance or military purposes, but even these should still be as strong as possible \footcite{Kerckhoffs1883}. ML based solutions may be Black Boxes for the two reasons mentioned above, but another could be the complexity of the neural network which supports it. The question then becomes whether CL systems should make use of such technology, if it is not possible to comprehend or in a transparent manner show what goes on inside of it. Considering the weight which CL systems in the future may have, in both a human and legal manner, it may simply not be advisable to use them on this basis alone.  Outside the reasoning above, it could also be due to the security worries which a system you do not know or understand can have.

\subsection{Authentication}


When authentication\footnote{For good taxonomy based overviews, see \cite{El-Hajj2017, Challa2018}.} is automated in a digital manner, it will be considered as being part of CL, regardless of whether it is incorporated or interpreted into the legal system.

But, authentication has a known amount of problems, such as proliferation of new technology, which does not solve its issues\footcite{Joos2022}, its practical constraints through circumvention\footcite{Blythe2013} or coercion, and lack of empirical research that really considers the population at large and not just students or paid individuals as the data source.

CL in authentication refers to the legal identification of the individual, either through a legal decision or as part of a process, and includes access to the state, banks and other companies which may constitute critical or important digital infrastructure in the lives of individuals. The conflict occurs when the CL is faulty, either deliberately or not, and leads to injustice or financial losses. Not unlike law in general, there is no quick solution, and CL must guarantee humans-in-the-loop to function in practice. 

In terms of cybersecurity, authentication may be an enabler for fraud through impersonation or integrity loss. Unlike past means of fraud, authentication in CL enables these actions to be done remotely and at a massive scale, which lowers the financial and practical costs and increases the risk of occurrence dramatically. This must be countered on a design or software level, but is rarely the focus of developers and users of these systems.


\section{Cybersecurity Legislation}\label{law}

While there is little concrete and hard law which directly regulates cybersecurity\footnote{This statement can be contested, on the basis that many jurisdictions will have overarching rules and legislation which may sound like it regulates it. The problem with a majority of these is that they leave the technical questions to guidance or worse, standards, without enforcement by professionals who actually understand these attributes. A further comparative legal analysis of this should be done elsewhere.}, we do have certain EU legislation which specifies elements of it. 


\subsection{The Cybersecurity Regulation and Future Legislation}

Currently, cybersecurity in the EU is regulated on a product to product type basis via guidance\footnote{If medical devices are used as an example, see our commentary on the role which cybersecurity guidance has in: \cite{Ludvigsen2022, Ludvigsen2021}. For other perspectives, see \cite{Banasinski2021, Chiara2022}.}, somewhat dedicated rules\footnote{Each type has its own Regulations and Directives. For a good example, see Regulation 2019/941 on risk-preparedness in the electricity sector and repealing Directive 2005/89/EC regarding critical infrastructure, [2019], L 158/1. Note that this is still not centralised hard law, but delegation. For general commentary, see \cite{Maglaras2018}.}, through the NIS directive\footnote{Directive 2016/1148, concerning measures for a high common level of security of network and information systems across the Union, [2016] L 194/1.} in a national and fragmented manner, strictly by the Cybersecurity Act\footnote{Through Regulation 2019/81 on ENISA (the European Union Agency for Cybersecurity) and on information and communications technology cybersecurity certification and repealing Regulation (EU)No 526/2013 (Cybersecurity Act), [2019] L 151/15.} which only applies to EU institutions, and coordinated in a semi-volunteered manner by ENISA via the Cybersecurity Act. On the sidelines, we do have standards and other measures that may be enforced on a contractual or national basis, but they come with the usual caveats. For CL, this means that until there is harder law specifying which techniques should and should not be used, it will exist in a grey area. Yet again, CL will be treated like other types of software and systems, even if there should perhaps be specialised rules to accommodate the increased risks caused by failure, not unlike the cybersecurity requirements that exist for critical infrastructure like telecommunication\footnote{See \cite{Szczepaniuk2022} for analysis on human factors in it.}.

\subsubsection{Cybersecurity Resilience Act}\label{CAA}

Earlier in 2022, the Commission called for evidence for a future Cybersecurity Resilience Act\footnote{See \cite{Ludvigsen2022b}.}. We provided commentary for this, but there are some further considerations when discussing CL. Resilience is usually defined as a system that enables detection, tolerance and recovery from issues\footnote{Defined more narrowly and precisely in \cite[251 - 252]{Anderson2020}.}. The problem with CL, is that there must be adequate legislation and redundancies and mechanisms, not just technical or engineering based solutions.  Real resilience in CL is thus only attained when we have had to time study the failures and issues of past systems, such as the SyRi case\footcite{Rachovitsa2022}, and this is traditionally how we improve both safety and security. Resilience in CL must therefore be a matter of increased redundancy, in the form of subsidiary systems or humans which can take over roles of the CL system in an emergency, strict logic and design which focuses heavily on preventing adversarial or non-adversarial failures through standards or existing research on issues and problems with the hardware or software used, and recovery mechanisms which allow the CL system to keep functioning either partially or fully after disruption of any kind. The latter should be both recovery from actions of the system, so that the resources (time, financial) are not wasted, but also in a classic sense, where it can recover from attacks which bring down the or partially hijacks the system. There is currently no interest in creating these in the future Cybersecurity Resilience Act, but there may be another approach to this problem.

\subsection{Artificial Intelligence Act}

CL will become part of the European product legislation world via the proposed AI Act\footnote{Proposal for a Regulation of the European Parliament and of the Council laying Down Harmonised Rules on Artificial Intelligence (Artificial Intelligence Act) and Amending Certain Union Legislative Acts, COM/2021/206.}. This means that both its cybersecurity\footnote{Art 15.} and its mechanisms as artificial intelligence\footnote{Through what kind of risk it constitutes, which a lot of the time will be high, see Art 6(2) and Annex III. CL can fit many of the categories specified in Annex III, which means it will mostly be considered High Risk AI.} will be the subject of regulation. How much and and of which kind remains to be seen, as the Act has not been finalised yet. Irrespective of the progress of the Act, we can draw some general considerations which CL bring. The proposed AI Act allows the EU to regulate CL in a practical manner, and since the type depends on the area which the CL system functions in, it can either draw rules from there\footnote{Various product legislation will work alongside the AI Act, see Art 6(2) and  Annex II.}, or perhaps be granted new guidance which it must follow under the supervision of National Competent Authorities (authorities). Sadly, if the same type of loose enforcement seen in other types of product legislation continues, there is a risk of easy circumvention or mere loose slaps on the wrists for private or public providers of CL, which may not be very beneficial for the subjects which will be affected by poorly secured CL systems.

\subsubsection{Cybersecurity and Resilience}

Let us \emph{firstly} take a closer look at Article 15:

\say{\emph{1. High-risk AI systems shall be designed and developed in such a way that they achieve, in the light of their intended purpose, an appropriate level of accuracy, robustness and cybersecurity, and perform consistently in those respects throughout their lifecycle. ...}}

The AI Act considers robustness and cybersecurity here, and the wording is much clearer and closer to the expectations of those that build these systems. The Article continues with:

\say{\emph{... 3. High-risk AI systems shall be resilient as regards errors, faults or inconsistencies that may occur within the system or the environment in which the system operates, in
particular due to their interaction with natural persons or other systems. The robustness of high-risk AI systems may be achieved through technical redundancy solutions, which may include backup or fail-safe plans. High-risk AI systems that continue to learn after being placed on the market or put
into service shall be developed in such a way to ensure that possibly biased outputs due to outputs used as an input for future operations (‘feedback loops’) are duly
addressed with appropriate mitigation measures. ...}}

As CL will mostly be considered High-risk, these rules will apply directly. The definition of resilience here excludes what is considered robustness, but in a combined understanding, the Article covers what resilience traditionally is. All in all, these are clear and well aligned with best practice, and even include considerations which we did \hyperref[ml]{earlier} regarding machine-learning. The main problem then becomes enforcement, and whether those that use or develop the systems will actually adhere to the rules.

\subsubsection{Human Oversight}\label{art14}

\emph{Secondarily}, the AI Act also considers human oversight in Article 14, which from a literal reading does come very close to the kind of rules needed to regulate CL adequately:

\say{\emph{1. High-risk AI systems shall be designed and developed in such a way, including with appropriate human-machine interface tools, that they can be effectively overseen by natural persons during the period in which the AI system is in use.}}

\say{\emph{2. Human oversight shall aim at preventing or minimising the risks to health, safety or fundamental rights that may emerge when a high-risk AI system is used in accordance with its intended purpose or under conditions of reasonably foreseeable
misuse, in particular when such risks persist notwithstanding the application of other requirements set out in this Chapter. ...}}

Article 14(1) and 14(2) entail effective oversight, a rather strong and non-negotiable position, which warrants transparency and ease of use, perhaps an issue for CL systems which make use of machine-learning. Minimizing risk via human oversight is very sound, especially regarding misuse\footnote{Which could be adversarial failures.}. Continuing this, Article 14(3) requires built and implemented or identified human oversight by the provider. This is a logical extension of the ideas above.  Article 14(4) specifies exactly what the human oversight should be capable off, effectively professional or system requirements, including full understanding, tendency and bias awareness, correct interpretation, knowing when to disregard the AI, and being able to intervene or stop the system. 

Overall, Article 14 has taken all of the best elements of how humans in-the-loop should be implemented, and expressed it in a fairly comprehensive manner. Regarding CL, the complexity will exist in how the system handles and understands the law, and the Article does not resolve this issue, but definitely leads the way. However, like above, we are left without opportunities for sanctioning, and there are no concrete definitions to find here or in the Annexes as to what exact behaviour we are looking for, as this will be fairly complicated when actually defined narrowly in internal rules or guidance. This is unlike cybersecurity, where techniques and concepts like requiring encryption and air gaps are fairly technical, but very possible to require in hard law.

\subsubsection{Enforcement}

The enforcement structure is quite important to consider when discussing CL, as there will be a certain overlap between it and AI. Ideally, the AI Act ends up in a form or is accompanied by additional regulation which enables it to handle the specific consequences of CL as AI rather well\footnote{If Annex I stays in its current shape, all CL will be considered AI unless they provide assistance, but even some of these may pass.}, but in its current form, this is not the case. However, the AI Act is equipped with obligations and requirements, and some means to enforce them, albeit not as harshly or directly as many had wanted\footnote{See critical analysis in \cite{Raposo2022} for more.}. The obligations are found in Articles 16 - 24 and 61 - 62, specifically for providers\footnote{Additional obligations for importers and distributors exist as well, and uniquely, some obligations for users, which is extremely important for CL systems. This is because unintended use or unexpected consequences, in litigation, may hinge on who caused it or additionally, product liability rules.}. There are additional enforcement measures in Articles 63 - 68, which may partially rely on the obligations above. Like other product legislation, the authority is (usually) not the body which will technically test whether the product conforms with the rules. Instead, this is done by Notified Bodies\footnote{Art 33, but be aware of the role which subsidiaries will play, see Art 34 for this.}, which will likely be private organisations, again a parallel to existing systems in, e.g., the medical device world in the EU. Secondarily, you have the aforementioned National Competent Authorities\footnote{Art 59.} who also act as Notifying Authorities\footnote{Art 30.}. Most of the duty of care or fulfillment of responsibilities, and even reporting, are put on the shoulders of the providers. In a CL context, this can be rather dangerous, as the 15 day notice in Article 62, by virtue of how long it is, can cause massive damage to individuals wrongfully decided on or assisted by the CL system. The enforcement mechanisms themselves rely on Article 64, which should give the authorities access to pretty much everything regarding the AI, but for them to independently step in and investigate, there must be a risk at a national level\footnote{Art 65(2).}. CL will not fulfill the requirements to be considered a risk on this stage, so Article 67 or 68 must be used instead. Using the same evaluation as in Article 65, the authority can force the provider of AI to withdraw and, if possible, repair and prevent the risk which the AI poses\footnote{Art 67(1).}. Worth noting is the definition of risk or breach of obligations in Article 67(1):

\say{\emph{... it presents a risk to the health or safety of persons, to the compliance with obligations under Union or national law intended to protect fundamental rights
or to other aspects of public interest protection ...}}

For CL, the latter part is intrinsically relevant, and if literally and loyally followed, could be the supporting stone which CL needs in its proper deployment. The problem then becomes the authorities themselves, \emph{whether they will realise and act on this in time} and other general enforcement issues. The first is (uniquely) answered in Article 59(4), but this can be abused or deliberately not followed, causing the authority to become inactive or at least barely functioning. For the second part, the wording in Article 65(2) is ambiguous, and can be interpreted in a variety of ways. If we follow the logic and ideas from existing product legislation\footnote{And since there is nothing explicitly stating when, where, and how market surveillance could ideally be done on AI in the AI Act.}, it must rely on self-reporting\footnote{Art 61 and 62.} or information from other national authorities, or worse, through disclosures from journalists or researchers in an informal manner. Lastly, the AI Act delegates the means which the authorities can withdraw or otherwise sanction the providers to the Member States, which sadly gives little certainty going forward. In a CL context, this means that ongoing injustice or financial damage could take a long time be discovered and resolved, and in this sense, an \emph{ex ante} review mechanism of the AI with CL features would be much more adequate. 

\section{Recommendations} \label{recommendations}

If we consider the clash that CL and cybersecurity bring, we can combine it with common sense arguments from both safety and security engineering, and the ideas found in the AI Act, forming three recommendations:

\begin{enumerate}
    \item CL systems must accommodate the cybersecurity threats which they can include, at a design, deployment and post-deployment level (life-cycle), considering all levels of \hyperref[sectionweakness]{weakness}. 
    \item CL systems should only be used when adequate, as it is not a silver bullet for every problem it is applied to. Adequate analysis and predictions must be independently made, with the caveats which this brings through likelihood and assumptions.
    \item Humans must still play a role in CL, as they represent plasticity, adaptability and arbitration, values which no CL system can possess. This is the only way which the systems can handle unexpected occurrences, systemic injustice and other difficulties. 
\end{enumerate}

\subsection{Comments}

While these are more general and will apply to a general comprehension of CL and cybersecurity in general, details and the understanding of each must be commented on briefly.

\paragraph{First recommendation.} 

This concerns itself with defences and mitigation measures towards adversarial threats. Many of these techniques or measures may follow from existing best practice, rules or certifications, but for CL, the consequences may be as severe as when cybersecurity is breached in medical devices\footnote{We discuss the potential consequences and attacks on medical devices like surgical robots in \cite{Ludvigsen2022}.} or other areas which can damage individuals or human rights. What follows from this, is taking it gravely seriously, not only because of the consequences, but also the procedural risk that comes with it, as failures will warrant lawsuits of various kinds depending on the jurisdiction. On the other hand, if deployed outside of liberal democratic states, these considerations can be removed and CL can be used with impunity, as is seen in China\footcite{Liang2022}, even if this is not advisable. Note that wording is soft, accommodate only indicates what is reasonable to expect, and does not mean that it cannot be merely resilient to the failures. As we noted \hyperref[CAA]{earlier}, resilience includes recovery, making a broader and inclusive wording more fitting.

\paragraph{Second recommendation.}

There are times where machine-learning\footcite{Chien2022} or other types of CL may not be adequate. \emph{Firstly}, because manual decision-making or the existing judicial or public legal system is enough to handle to problem, or \emph{secondarily}, because the proportionality or the consequences on rights or finances on individuals may weigh greater than implementing the system. Regardless of this, choosing to use CL is a political decision, but this recommendation then warrants \emph{ex post} criticism and eventual dismantlement. There is little appreciation of the conservative aspect of non-choice or denial of certain uses of technology in existing rules, but this is traditional and well known in safety engineering. There will simply be times where safety risks outweigh the eventual advantages a system may bring, which means it should not be used, regardless of the financial or personal interests in it. This should in particular apply to CL because of its potentially systemic damage capabilities.

\paragraph{Third Recommendation.}

Mandating humans in systems is not in any way new, but we ask for it for a more traditional reason: The qualities which only humans have\footnote{But it may have negative consequences, if used in a manner to circumvent or otherwise abuse the good ideas behind human oversight, see \cite{Green2022}.}. This is, again, a reference to the role which humans have in safety engineering, where our ability to adapt in emergencies make us invaluable\footcite[101 - 102]{Leveson1995}. Of course, the caveat to this is where adversarial behaviour or sludging\footnote{Much great literature exists on this, see \cite{Paul2022}, for a recent overview and example.} causes us to not react appropriately. Still, humans must act as both emergency measures and redundancies, the latter specifically referring to humans stepping in and overtaking the role which CL had so far. This could be in decision-making or automated assistance, both fields where humans used to sit, which means we still have experience and infrastructure to provide it, albeit much slower than what CL can provide. \hyperref[art14]{Article 14} in EU's proposed AI Act is a great example of how such a rule could be designed.


\section{Future Work}

Continuous in-depth analysis of existing CL systems, and acknowledging what constitutes CL to include as much as possible is, in our view, warranted. This should include interdisciplinary aspects, as understanding one side or another is not sufficient to actually comprehend the system as a whole, consequences, constraints and all. 

The role which economics play in both enforcement or lobbying by powerful corporations or other states should be considered regarding CL too. Essentially, economic studies which critically consider the potential loopholes and adversarial actions that powerful actors can use against CL systems, is wanted on the basis of this paper, perhaps extending existing economics of security ideas.

Based on our very informal recommendations, expanded analysis on what constitutes the right situation to deploy CL, a condensed analysis of which threats for which types of CL exist, and the necessary human roles needed in CL would be adequate.


And finally, fusing CL research with existing but much older digitalised law\footnote{Such as the works by Jon Bing.} and legal informatics is suitable, as both fields have common agendas and perceptions of digital law, and would benefit from the expertise and alternative starting points which each bring.

\section{Conclusion}

As it is clear, CL presents a situation where law gains both the advantages and disadvantages of the other discipline which inspired its existence\footnote{Not unlike the issues which philosophy or political directions has caused of systemic damage to law in the past. An example of this could be ``communist law'', see e.g., \cite{Priban2009}.}. In this paper, we firstly made a conceptualisation of cybersecurity in CL specifically, which consists of \hyperref[sectionweakness]{five levels}. Now that CL absorbs cybersecurity weaknesses which occur in each layer, we show that there exists mitigation measures for the problems. But we know from safety and security engineering, that an ongoing arms race is and will be needed going forward, a sort of constant ongoing development of defences and considerations, which attempts to mitigate or lessen the damage that attacks can cause. Building these into CL will be vital, just as it is with cybersecurity everywhere else.

We find that CL is vulnerable to the same issues which both AI and neural networks face, if the systems make use of the technology\footnote{This may seem rather redundant to mention, but these tend to be overlooked in favour of what advantages the systems may bring. In this sense, making sure that all aspects of new technology are discussed and understood is a core role of researchers and the public alike.}. The logic and practical implementation in which CL is implemented, presents a danger which very few other legal disciplines contain. Thereafter, we show how fragmented and rather softly cybersecurity is regulated in the EU, which in relation to CL, presents practical and real problems. But, we then analyse the AI Act, and it provides sound measures known from elsewhere regarding cybersecurity and human oversight. Regardless of issues with enforcement, this illustrates promise and understanding for a future where CL will play a central role in the legal system and in public consciousness in general. Clearly, the AI Act here combines CL (as AI) and cybersecurity, and regulates them together in a \hyperref[law]{legal sense}.  Finally, we provide \hyperref[recommendations]{three recommendations} for CL related to cybersecurity, which summarised are: Accommodation of threats, adequate use of CL, maintenance of the human role.

These are general, but from a EU perspective, the AI Act already does a good job in implementing them all, giving us hope for a future where CL and cybersecurity can safely melt together.


\end{document}